\documentclass{optica-article}

\journal{opticajournal} 

\articletype{Research Article}
\usepackage{rsfso}
\usepackage{hyperref}

\usepackage{graphicx}
\usepackage{dcolumn}
\usepackage{bm}
\usepackage{url}

\usepackage{mathtools}
\usepackage{braket}
\usepackage{lineno}

\newcommand{\unit}[1]{\ \mathrm{#1}}
\newcommand{\sgn}{\text{sgn}}

\begin{document}

\title{
\textit{In situ} substrate birefringence characterization in gravitational wave detectors using a heterodyne polarimetry method
}

\author{Satoshi Tanioka,\authormark{1,*} 
Terri Pearce,\authormark{1}
Yuta Michimura,\authormark{2,3}
Kazuhiro Agatsuma,\authormark{4}
Martin van Beuzekom,\authormark{5}
Alberto Vecchio, \authormark{4}
Stephen Webster, \authormark{6}
Matteo Leonardi, \authormark{7}
and Keiko Kokeyama\authormark{1}}

\address{\authormark{1}School of Physics and Astronomy, Cardiff University, Cardiff CF24 3AA, United Kingdom\\
\authormark{2}Research Center for the Early Universe (RESCEU) , Graduate School of Science, The University of Tokyo, 7-3-1 Hongo, Bunkyo, Tokyo 113-0033, Japan \\
\authormark{3}Kavli Institute for the Physics and Mathematics of the Universe (Kavli IPMU) , WPI, UTIAS, The University of Tokyo, 5-1-5 Kashiwanoha, Kashiwa, Chiba 277-8568, Japan \\
\authormark{4}Institute for Gravitational Wave Astronomy, School of Physics and Astronomy, University of Birmingham, Birmingham B15 2TT, United Kingdom \\
\authormark{5}Nikhef, National Institute for Subatomic Physics, Science Park 105, 1098 XG Amsterdam, Netherlands \\
\authormark{6}SUPA, University of Glasgow, Glasgow G12 8QQ, United Kingdom\\
\authormark{7}Dipartimento di Fisica, Universit` a di Trento, 38123 Povo, Trento, Italy \\
%
}
\email{\authormark{*}taniokas@cardiff.ac.uk} 


\begin{abstract*} 

High-quality test mass substrates play essential roles in laser interferometric gravitational wave detectors.
Inhomogeneous birefringence distribution in test mass substrates, however, can degrade the sensitivity of the detector by introducing the optical loss and disturbing the interferometer controls.
In this paper, we present a heterodyne polarimetry method that enables \textit{in situ} birefringence characterizations, hence diagnosing the gravitational wave interferometer.
We experimentally demonstrate the proposed method with a tabletop setup.
We also discuss its applicability to current and future gravitational wave detectors and the detectable limit.

\end{abstract*}

\section{Introduction}

Ground-based laser interferometric gravitational wave detectors (GWDs) are based on dual-recycled Fabry-Perot Michelson interferometers to achieve sufficient sensitivity for direct detection of GWs \cite{Aso2013, Aasi2015, Acernese2015}.
Those GWDs have opened a new window on the Universe by enabling direct detection of gravitational waves (GWs) \cite{Abbott2016}.
The multi-messenger observation of GW170817, especially, provided fruitful insights into the origin of the heavy elements in the universe \cite{Abbott2017}.
Future GWDs, such as Cosmic Explorer and the Einstein Telescope, will enhance the scientific impact of gravitational wave astronomy by increasing the sensitivity by one order of magnitude compared to current detectors that will enable us to observe the binary black hole merger up to $z\sim20$ \cite{Punturo2010, Hall2021}.

High-quality test masses play a crucial role in both improving the performance of current GWDs and in the realization of the next-generation GWDs \cite{Degallaix2019}.
One of the important properties to be assessed in such test masses is the birefringence distribution.
Recent studies in KAGRA have showed that inhomogeneous birefringence in crystalline sapphire input test masses (ITMs) degrade the performance of the detector \cite{Somiya2019, Wang2024}.
Such inhomogeneous birefringence distribution in an ITM is known to affect the detector performance as it can degrade the power-recycling gain (PRG), resulting in reduced circulating power \cite{Winkler1994}.
In addition to the PRG reduction, the inhomogeneous birefringence of an ITM can affect the contrast of the interferometer, which can cause worse technical noise coupling such as the laser frequency noise,  \cite{Aasi2015, Krüger2016}.

While a tabletop setup can characterize the test mass birefringence distribution, the actual birefringence distribution can be altered by the stress induced by a suspension or by cooling down the test mass to cryogenic temperature.
In particular, next-generation GWDs will employ a few hundred kilogram test masses, which may cause substantial stress to induce inhomogeneous birefringence \cite{Punturo2010, Hall2021, Hamedan2023}.
In addition, their test masses are thicker than the current test masses which can therefore enhance the phase retardation as shown in Eq. (\ref{eq:retardation}).
Even if the test mass birefringence distribution is pre-characterized by a tabletop setup, the installation process to the GWD can introduce unexpected birefringence due to strain applied to the test masses whose effect has not been rigorously evaluated.
Test mass birefringence may also vary during long term observation, a possibility that has also not been systematically tested.
For instance, operating the GWD under high beam power may thermally induce birefringence in the test mass \cite{Winkler1994}.
\textit{In situ} birefringence characterization will play an important role in understanding the performance of the GWDs and determining commissioning strategies by providing birefringence distributions.
Such birefringence characterization can also act as a useful tool to study the \textit{in situ} material properties.

In tabletop experiments, the test mass birefringence can be characterized by rotating either the input beam polarization or the sample mirror itself \cite{Zeidler2023, Wang2024}.
However, neither can be rotated in GWDs.
Therefore, in order to realize the \textit{in situ} birefringence characterization, it is necessary to develop a method which does not require rotation of the input beam polarization or the test mass.

In this paper, we propose and demonstrate a heterodyne polarimetry approach to enable in-situ test mass birefringence characterization in a gravitational wave detector.
To validate this concept, we have developed a tabletop setup which shows reasonable repeatability and sensitivity.
We also discuss a possible configuration for GWDs and detectable limit.

\section{Theory and method} \label{sec:method}

In this section, we derive the definition of substrate birefringence.
Then, we will show the birefringence extraction method based on heterodyne wavefront sensing.

\subsection{Definition}

\begin{figure}[htbp]
\centering\includegraphics[width=10cm]{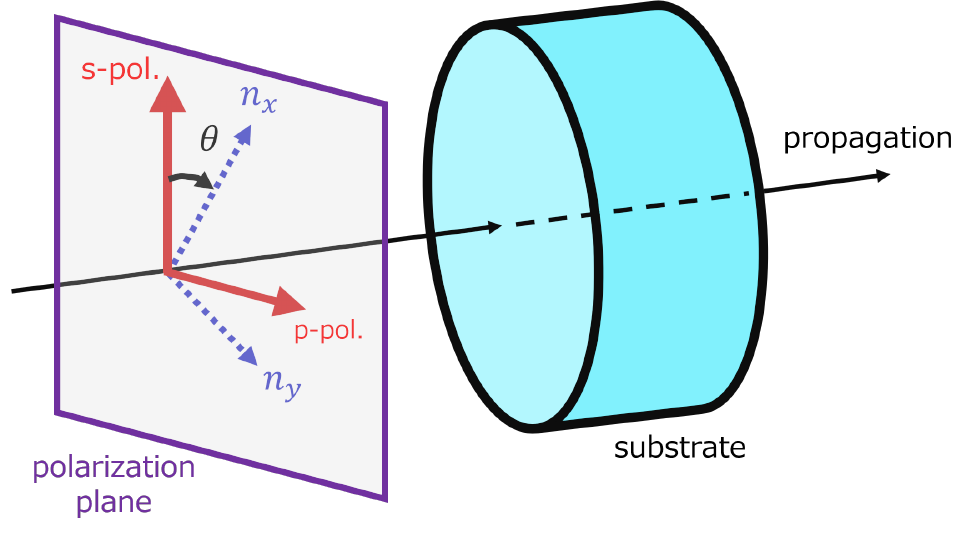}
\caption{
Schematic of the definition.
The mirror substrate has two orthogonal refractive index axes, $n_{x}$ and $n_{y}$, which are rotated by $\theta$ from s- and p-polarization axes, respectively.
}
\label{fig:definition}
\end{figure}

We define parameters and geometry to characterize the birefringence distribution as shown in Fig. \ref{fig:definition}.
In the same manner as the previous work, we set s- and p-polarization axes in the vertical and horizontal directions with respect to the laboratory frame \cite{Wang2024, Tanioka2025}.
The polarization state of the light field can be expressed as
\begin{align}
    \ket{E} =
    \begin{pmatrix}
    E^{\mathrm{s}} \\
    E^{\mathrm{p}}\\
    \end{pmatrix},
\end{align}
where $E^{\mathrm{s}}$ and $E^{\mathrm{p}}$ are the electric fields of the s- and p-polarized fields, respectively.

The property of substrate birefringence can be described by refractive index difference, $\Delta n$, between two distinctive orthogonal axes, $n_x$ and $n_y$.
These two orthogonal refractive index axes are respectively called ordinary and extraordinary axes.
Here, we assume that these axes are misaligned by an angle of $\theta$ with respect to the polarization axes as shown in Fig. \ref{fig:definition}.

The polarization state of the light field can be modified when it passes through the birefringent substrate.
We define the following common and differential phase changes as employed in  the previous works \cite{Wang2024, Tanioka2025}
\begin{align}
    \alpha_+ \coloneqq \frac{\alpha_x+\alpha_y}{2} = \frac{\pi d}{\lambda}\left(n_x+n_y\right), \\
    \alpha_- \coloneqq \frac{\alpha_x-\alpha_y}{2} = \frac{\pi d}{\lambda}\left(n_x-n_y\right), \label{eq:retardation}
\end{align}
where $d$ is the thickness of the substrate and $\lambda$ is the laser wavelength.
Then, the Jones matrix of the birefringent sample, $\hat{M}$, can be expressed as
\begin{align}
    \hat{M}
    &= \mathrm{e}^{i\alpha_+}
    \begin{pmatrix}
        \cos\alpha_- + i\sin\alpha_-\cos2\theta & i\sin\alpha_-\sin2\theta \\
        i\sin\alpha_-\sin2\theta & \cos\alpha_- - i\sin\alpha_-\cos2\theta \\
    \end{pmatrix}.
    \label{eq:mirror}
\end{align}
In this convention, the range of the orientation and differential phase retardation are $-45^{\circ}\leq\theta\leq45^{\circ}$ and $-\pi/2\leq\alpha_-\leq\pi/2$, respectively.

The electric field of the s-polarized input beam can be described as
\begin{align}
    \ket{E_{\mathrm{in}}} &=
    \begin{pmatrix}
    E_{\mathrm{in}} \\
    0 \\
    \end{pmatrix},
\end{align}
where $E_{\mathrm{in}}\coloneqq  |E_0|\mathrm{e}^{i\Omega t}$, $|E_0|$ is the amplitude of the field, and $\Omega$ is the laser angular frequency.
When the s-polarized beam passes through the birefringent sample, the transmitted light field becomes
\begin{align}
    \ket{E_{\mathrm{out}}} \coloneqq
    \begin{pmatrix}
    E^{\mathrm{s}}_{\mathrm{out}} \\
    E^{\mathrm{p}}_{\mathrm{out}}\\
    \end{pmatrix}
    = \hat{M}\ket{E_{\mathrm{in}}} = E_{\mathrm{in}}\mathrm{e}^{i\alpha_+}
    \begin{pmatrix}
        \cos\alpha_- + i\sin\alpha_-\cos2\theta \\
        i\sin\alpha_-\sin2\theta \\
    \end{pmatrix}.
    \label{eq:aftersample}
\end{align}
The above equation indicates the input beam field is scattered into orthogonal polarization, whose amplitude can be described as $|E^{\mathrm{p}}_{\mathrm{out}}|=|\sin\alpha_-\sin2\theta|$.
The square of this term, i.e., $|E^{\mathrm{p}}_{\mathrm{out}}|^2=\sin^2\alpha_-\sin^22\theta$ can be regarded as an optical loss induced by the birefringence.
In addition to the common phase change $\alpha_+$, s- and p-polarized beams experience phase changes depending on the arguments of their complex fields.
The accumulated phase of each polarization can be expressed as
\begin{align}
    &\arg\left(E^{\mathrm{s}}_{\mathrm{out}}\right) = \arg\left(\cos\alpha_- + i\sin\alpha_-\cos2\theta\right) + \alpha_+ = \arctan\left(\tan\alpha_-\cos2\theta\right) + \alpha_+,  \label{eq:sphase}\\
    &\arg\left(E^{\mathrm{p}}_{\mathrm{out}}\right) = \arg\left(i\sin\alpha_-\sin2\theta\right) + \alpha_+ = \frac{\pi}{2}\sgn\left(\sin\alpha_-\sin2\theta\right) + \alpha_+, \label{eq:pphase}
\end{align}
where $\sgn(\sin\alpha_-\sin2\theta)$ denotes the sign of $\sin\alpha_-\sin2\theta$,
\begin{align}
    \sgn(\sin\alpha_-\sin2\theta) = \begin{cases}
        +1\quad(\sin\alpha_-\sin2\theta>0), \\
        -1\quad(\sin\alpha_-\sin2\theta<0).
    \end{cases}
\end{align}


\subsection{RF heterodyne detection}

To obtain birefringence distribution maps, we employ the RF heterodyne wavefront scheme as demonstrated in Refs \cite{Goda2004, Agatsuma2019}.
In this scheme, the beatnote between the test beam which passes through the birefringent substrate and the frequency shifted reference beam is detected by an RF photodiode (PD).
Here we define the laser field on the PD which experiences birefringence material and the reference field as $E_{\mathrm{PD}}^j$ and $E_{\mathrm{ref}}^j$, respectively.
Here, $j$ denotes the polarization, i.e., $j\in(\mathrm{s}, \mathrm{p})$.
When the laser angular frequency of the reference beam is shifted by $\omega$, the beatnote between the test and reference beams at PDs become
\begin{align}
    |E_{\mathrm{PD}}^{j}|\mathrm{e}^{-i\Omega t}\mathrm{e}^{-i\phi_{\mathrm{test}}^j}|E_{\mathrm{ref}}|\mathrm{e}^{i(\Omega + \omega) t}\mathrm{e}^{i\phi_{\mathrm{ref}}^j} + |E_{\mathrm{PD}}^{j}|\mathrm{e}^{i\Omega t}\mathrm{e}^{i\phi_{\mathrm{test}}^j}|E_{\mathrm{ref}}^{j}|\mathrm{e}^{-i(\Omega + \omega) t}\mathrm{e}^{-i\phi_{\mathrm{ref}}^j},
\end{align}
where $\phi_{\mathrm{test}}$ is the phase retardation induced by birefringence in the sample, and $\phi_{\mathrm{ref}}$ is the phase of the reference field.
Here, we defined the beatnote amplitude and relative phase as $A^j\coloneqq|E_{\mathrm{PD}}^{j}||E_{\mathrm{ref}}^{j}|$ and $\phi^j\coloneqq\phi_{\mathrm{ref}}^j-\phi^j_{\mathrm{test}}$.
The above equation becomes
\begin{align}
    A^j\cos\omega t\left(\mathrm{e}^{i\phi^j}+\mathrm{e}^{-i\phi^j}\right) + iA^j\sin\omega t\left(\mathrm{e}^{i\phi^j}-\mathrm{e}^{-i\phi^j}\right).
\end{align}
By demodulating this signal in-phase and quad-phase, we can obtain \cite{Goda2004}
\begin{align}
    V_{\mathrm{I}}^j &= A^j\cos\phi^j, \\
    V_{\mathrm{Q}}^j &= A^j\sin\phi^j.
\end{align}
Then, we can compute the beatnote amplitude and relative phase as
\begin{align}
    A^j &= \sqrt{(V^j_{\mathrm{I}})^2 + (V^j_{\mathrm{Q}})^2}, \\
    \phi^j &= \arg\left(V^j_{\mathrm{I}} + iV^j_{\mathrm{Q}}\right).
\end{align}


\subsection{Birefringence extractions}
\label{sec:extraction}
From Eqs. (\ref{eq:sphase}) and (\ref{eq:pphase}) and the previous subsection, the relative phase detected by the heterodyne measurement can be expressed as
\begin{align}
    \phi^j &= \phi_{\mathrm{ref}}^j - \arg\left(E^j_{\mathrm{out}}\right).
\end{align}
We define the phase difference between s- and p-polarizations as
\begin{align}
    \phi_{\mathrm{diff}} &\coloneqq \phi^{\mathrm{p}} - \phi^{\mathrm{s}}, \notag \\
    &= \arctan\left(\tan\alpha_-\cos2\theta\right) - \frac{\pi}{2}\sgn\left(\sin\alpha_-\sin2\theta\right) + \phi_{\mathrm{offset}},
    \label{eq:phase_diff}
\end{align}
where $\phi_{\mathrm{offset}}$ is an offset introduced by optical components, which are not shared in between the main beam and reference beam.

As long as the phase offset $\phi_{\mathrm{offset}}$ is pre-characterized and the sign of $\sin\alpha_-\sin2\theta$ is determined, we can extract $\arctan\left(\tan\alpha_-\cos2\theta\right)$ from Eq (\ref{eq:phase_diff}).
In order to extract the differential phase retardation, $\alpha_-$, we adopt the following approximation
\begin{align}
    (\sin\alpha_-\sin2\theta)^2 + \left\{\arctan(\tan\alpha_-\cos2\theta)\right\}^2 &\approx \alpha_-^2.
    \label{eq:approx2}
\end{align}
The first term in the above equation corresponds to the amplitude of the p-polarized beam, $|E^{\mathrm{p}}_{\mathrm{out}}|$ as shown in Eq. (\ref{eq:aftersample}).
By taking the square root of the above equation, we can compute the absolute value $|\alpha_-|$.

Once $|\alpha_-|$ is computed, the absolute value of the orientation $\theta$ can be derived by the following equation:
\begin{align}
    |\theta| &= \frac{1}{2}\arcsin\left(\frac{|E^{\mathrm{p}}_{\mathrm{out}}|}{|\sin\alpha_-|}\right).
    \label{eq:theta_approx}
\end{align}

\begin{figure}[htbp]
\centering\includegraphics[width=13.2cm]{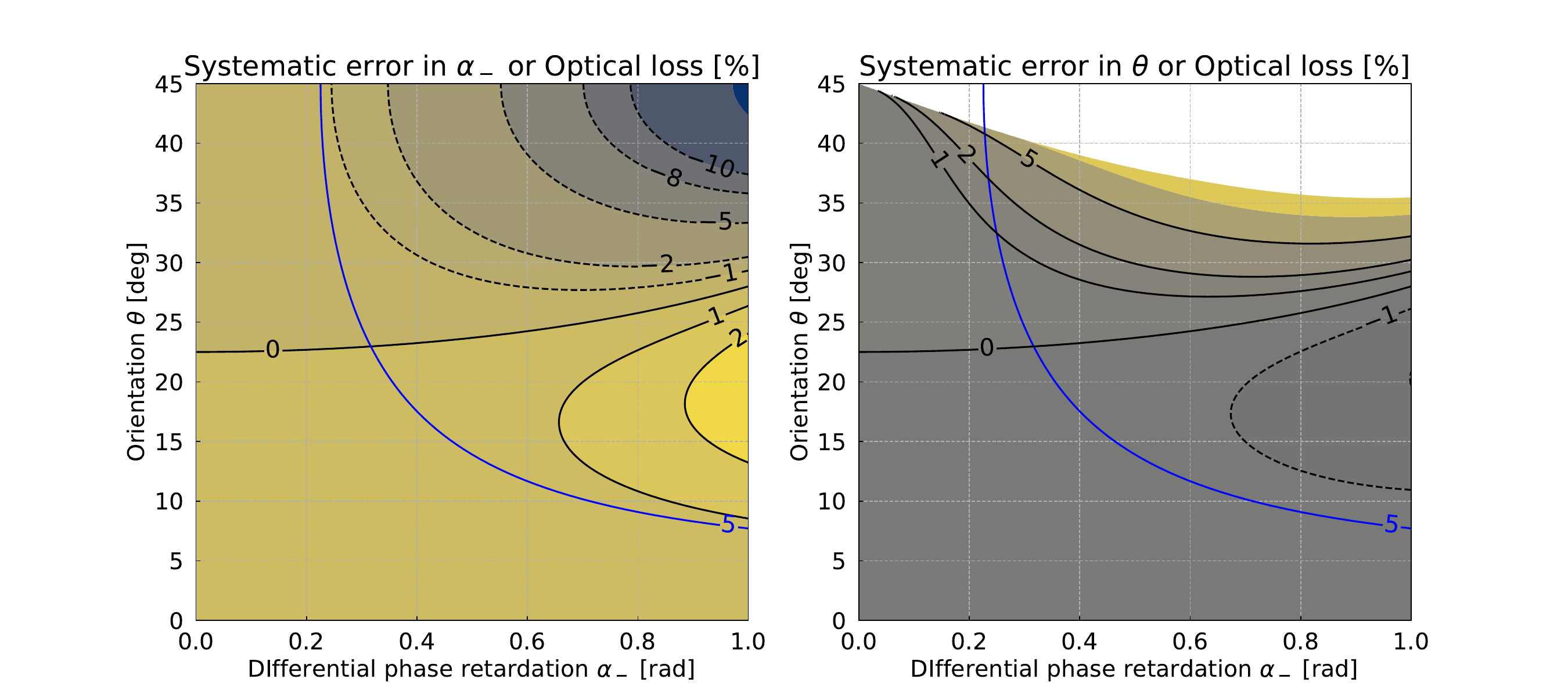}
\caption{
Contour maps of systematic errors in differential phase retardation and orientation.
Since there is a symmetry, only the first quadrant is plotted for each case.
The blue curve in each plot corresponds to the optical loss ($\sin^2\alpha_-\sin^22\theta$) of $5\%$.
}
\label{Fig:OpticalLoss}
\end{figure}

Fig. \ref{Fig:OpticalLoss} shows the systematic errors in $\alpha_-$ and $\theta$ computed by using the approximation Eq. (\ref{eq:approx2}).
The blue curves shown in those plots correspond to the birefringence optical loss of $\sin^2\alpha_-\sin^22\theta=5\%$.
For GWDs, the birefringence optical loss must be kept below a few percent, which will be discussed in later section.
Therefore, both $|\alpha_-|$ and $|\theta|$ need to be smaller than the values given this blue curve.
Under this condition, the systematic error of birefringence parameters, $|\alpha_-|$ and $|\theta|$, calculated using the approximation given by Eq. (\ref{eq:approx2}) is less than $\pm1\%$.
This accuracy is sufficient for diagnosing whether a GWD's performance is limited by birefringence-induced optical loss.
Where the axis orientation gets close to $45^\circ$, underestimated $|\alpha_-|$ causes $|E^{\mathrm{p}}_{\mathrm{out}}|/|\sin\alpha_-|>1$.
Therefore, $\arcsin(|E^{\mathrm{p}}_{\mathrm{out}}|/|\sin\alpha_-|)$ cannot be defined.
This results in the white region in orientation contour map.

In order to determine the value of $\sgn(\sin\alpha_-\sin2\theta)$, we use the following procedures.
First, we define $\tilde{\phi}_{\mathrm{diff}}\coloneqq\arctan\left(\tan\alpha_-\cos2\theta\right)-\pi/2\times\sgn(\sin\alpha_-\sin2\theta)$.
Then, the following conditions are satisfied depending on the sign of $\sin\alpha_-\sin2\theta$ as
\begin{align}
    \tilde{\phi}_{\mathrm{diff}} + \frac{\pi}{2} = \begin{cases}
        \arctan\left(\tan\alpha_-\cos2\theta\right)\quad(\sin\alpha_-\sin2\theta>0), \\
        \arctan\left(\tan\alpha_-\cos2\theta\right)+\pi\quad(\sin\alpha_-\sin2\theta<0).
    \end{cases}
\end{align}
From the definition of the inverse of the tangent function, the domain of definition of $\arctan\left(\tan\alpha_-\cos2\theta\right)$ is $-\pi/2 \leq\arctan\left(\tan\alpha_-\cos2\theta\right)\leq\pi/2$.
Therefore, we can divide into two cases:
\begin{align}   \left|\tilde{\phi}_{\mathrm{diff}} + \frac{\pi}{2}\right|<\frac{\pi}{2} &\Rightarrow \sin\alpha_-\sin2\theta>0, \\
\left|\tilde{\phi}_{\mathrm{diff}} + \frac{\pi}{2}\right|>\frac{\pi}{2} &\Rightarrow \sin\alpha_-\sin2\theta<0.
\end{align}
By doing the above comparison, $\sgn(\sin\alpha_-\sin2\theta)$ can be determined, hence $\arctan\left(\tan\alpha_-\cos2\theta\right)$ can be computed.
From computed $\arctan\left(\tan\alpha_-\cos2\theta\right)$ and Eqs. (\ref{eq:approx2}) and (\ref{eq:theta_approx}), the absolute values of birefringence parameters can be derived.
The sign of $\arctan\left(\tan\alpha_-\cos2\theta\right)$ gives the sign of $\alpha_-$.
Combining with the sign of $\sin\alpha_-\sin2\theta$, we can get the sign of $\theta$.
From the above procedures, we can obtain the birefringence parameters, $\alpha_-$ and $\theta$, including their signs.
This will allow us to monitor how the substrate birefringence vary over time and provide useful information to decide commissioning activities.

\section{Experimental setup}

\subsection{Optical layout}

\begin{figure}[htbp]
\centering\includegraphics[width=11cm]{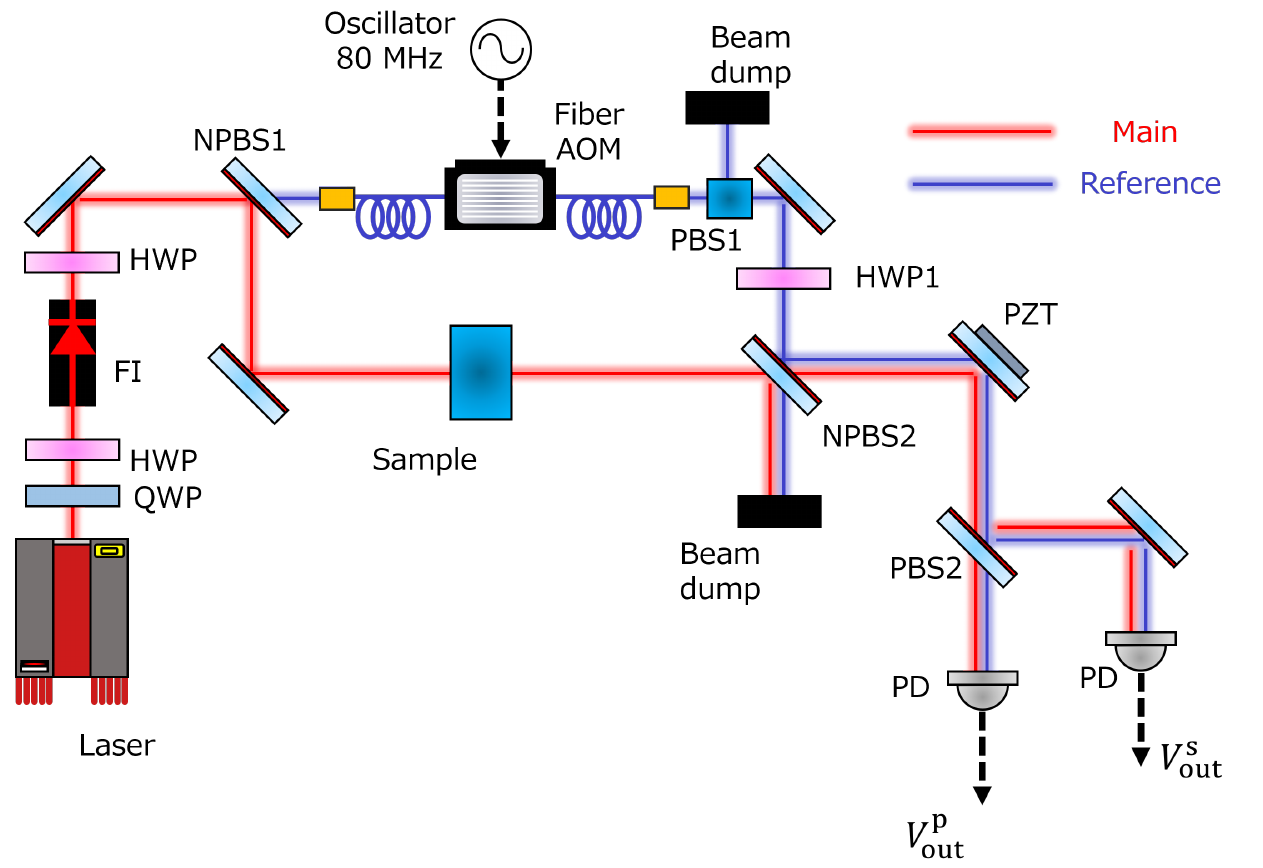}
\caption{
Schematic of the experimental setup.
The input beam is conditioned to be an s-polarized beam using polarizers.
A non-polarizing beamsplitter (NPBS1) divides the beam into two paths.
The frequency of the reference beam is shifted using a fiber AOM.
The test and the reference beams are combined by NPBS2, and then split into s- and p-polarized beams.
PBS1 is installed after the fiber output to remove the unwanted polarization component.
Then the reference beam power on each PD is adjusted by HWP1.
}
\label{Fig:layout}
\end{figure}

To demonstrate the heterodyne polarimetry, we developed a tabletop setup.
In particular, we adopted the phase camera setup which is utilized in Advanced Virgo to diagnose the interferometer \cite{Schaaf2016, Agatsuma2019}.
This setup enables us to probe the 2D birefringence distribution of the substrate.
Here, we refer to this apparatus as the polarization phase camera (PPC).

Fig. \ref{Fig:layout} shows the optical layout of the PPC setup.
A $1064\unit{nm}$ wavelength Mephisto laser is used as a laser source with a beam power of about $200\unit{mW}$.
The polarization of this input laser is adjusted to s-polarization by a combination of polarizers.
The input beam is split into two paths, reference and test beam paths, by using a non-polarizing beamsplitter (NPBS1).
In the reference beam path, a fiber acousto-optic modulator (AOM) is installed to generate a frequency shifted laser field for the heterodyne detection.
In our setup, the laser frequency is shifted by $+80\unit{MHz}$.
A birefringent sample is placed in the test beam path to characterize its birefringence distribution.
The polarization extinction ratio incident on the sample is kept better than $10^3:1$, i.e., the input p-polarization beam power is kept below $0.1\%$ compared to that of s-polarization beam.
After passing through the sample, the test beam is combined with the reference beam by another non-polarizing beamsplitter (NPBS2).
Then, those combined beams are split into s- and p-polarizations by a polarized beamsplitter (PBS), and detected by pin-hole photodetectors (PDs).
The PBS has the polarization extinction ratio better than $10^4:1$.
We use the same pin-hole PD as the previous work which has an aperture of $55\unit{\mu m}$ \cite{Agatsuma2019}.

Between NPBS2 and the PBS2, a two-axis piezoelectric transducer (PZT) scanner (S-334 by Physik Instrument) is installed.
By using this PZT scanner, both the test and reference beams are scanned so that we can get the their beatnote spatial distribution on PDs.
The distance between the scanner and each PD is set at $300\unit{mm}$.

The beam radii on PDs are set to $750\unit{\mu m}$ and $1550\unit{\mu m}$ for the test and the reference beams, respectively.
The reference beam is highly collimated thanks to the collimator at the fiber output.
The test beam is also highly collimated by using a pair of lenses.

For scanning the beams, the following Archimedean spiral pattern is chosen:
\begin{align}
    x(t) = R_{\mathrm{img}}\frac{t}{T_{\mathrm{img}}}\cos\left(2\pi f_{\mathrm{scan}}t\right), \\
    y(t) = R_{\mathrm{img}}\frac{t}{T_{\mathrm{img}}}\sin\left(2\pi f_{\mathrm{scan}}t\right).
\end{align}
$R_{\mathrm{img}}$, $T_{\mathrm{img}}$, and $f_{\mathrm{scan}}$ are the radius of image, the acquisition time for one image, and the spiral rotation frequency, respectively.
The spiral rotation frequency of $f_{\mathrm{scan}}=32\unit{Hz
}$ is employed.
This scanning pattern enables a smooth trajectory, thus smooth mirror movement.



For the demonstration purpose, a zero-order quarter-wave plate (QWP) is used as the sample because it has a known relative phase retardation of $|\alpha_-|=\pi/4$.
In addition, it has an indicator such that we can identify its fast or slow axis, i.e., the sign of $\alpha_-$ of the QWP.


\subsection{Data acquisition}

\begin{figure}[htbp]
\centering\includegraphics[width=11cm]{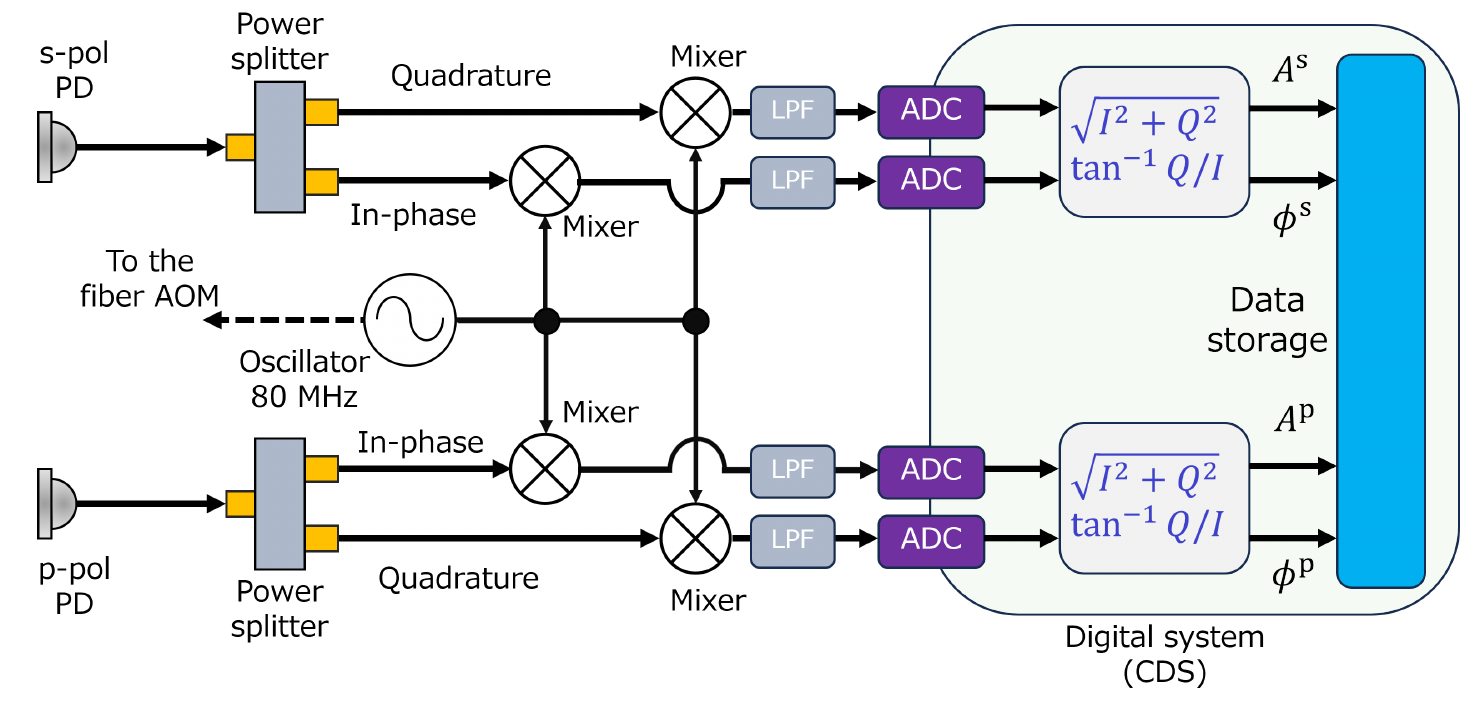}
\caption{
Schematic of the data acquisition system.
The beatnote signals are demodulated by analog RF circuits.
The demodulated signals are filtered by low-pass filters (LPFs), and then converted to digital signals by analog-to-digital converters (ADCs).
The post-processing and data storage are performed on the digital system.
}
\label{Fig:daq}
\end{figure}

Fig. \ref{Fig:daq} shows the sequence of data acquisition.
The RF beatnote signal from each PD is split by a power splitter, and demodulated in-phase and quadrature by mixers, respectively.
The demodulated analog signals are low-passed by low-pass filters (LPFs) and then converted to digital signals by by 16 bit analog-to-digital converters (ADCs).
The post-processing to extract the beatnote amplitude and phase is done by a digital system called the control digital system (CDS), which is used in Advanced LIGO \cite{Aasi2015}.
The computed beatnote amplitude, $A^j$, and phase, $\phi^j$, are recorded for each measurement.
By using these recorded data, the post-processing is performed to extract birefringence maps as described in Sec. \ref{sec:method}.

The ADC used in this experiment has $16384$ sampling per second.
Maps are consisting of $128\times128$ points, hence $16384$ pixels.
We set the scanning time to be $1\unit{s}$, yielding 1 sample per each pixel.
Note that once a FPGA based data acquisition system is deployed as shown in \cite{Agatsuma2019}, samples per pixel can be enhanced to $16384$ with $500\unit{MS/s}$ sampling rate.



\section{Results}

\subsection{Calibration}

\subsubsection{Amplitude}

In order to estimate the birefringence using Eq. (\ref{eq:approx2}), we need to calculate $\sin^2\alpha_-\sin2\theta$.
This term can be computed by calibrating the measured beatnote amplitudes.

Between the sample and the PDs, optical components are placed in order to constitute the PPC such as NPBS2.
These optical components introduce optical attenuation unequally to s- and p-polarizations.
This can be expressed as
\begin{align}
    \begin{pmatrix}
    E^{\mathrm{s}}_{\mathrm{PD}} \\
    E^{\mathrm{p}}_{\mathrm{PD}}\\
    \end{pmatrix} &=
    \begin{pmatrix}
        \sqrt{T^{\mathrm{s}}} & 0 \\
        0 & \sqrt{T^{\mathrm{p}}}
    \end{pmatrix}
    \begin{pmatrix}
    E^{\mathrm{s}}_{\mathrm{out}} \\
    E^{\mathrm{p}}_{\mathrm{out}}\\
    \end{pmatrix},
\end{align}
where $T^j$ represent the power transmissivities.
Each transmissivity was measured by a power meter (S132C from Thorlabs) and determined as $T^{\mathrm{s}}=0.375$ and $T^{\mathrm{p}}=0.545$, respectively.
Note that the asymmetry in the transmissivity mostly originates from NPBS2.

Here we consider the following calculation
\begin{align}
    \frac{\left(A^{\mathrm{p}}\right)^2}{\left(A^{\mathrm{s}}\right)^2 + \left(A^{\mathrm{p}}\right)^2}
    &= \frac{\sin^2\alpha_-\sin^22\theta \times T^{\mathrm{p}}P_{\mathrm{in}}P_{\mathrm{ref}}^{\mathrm{p}}}{\left(1-\sin^2\alpha_-\sin^22\theta\right)\times T^{\mathrm{s}}P_{\mathrm{in}}P_{\mathrm{ref}}^{\mathrm{s}} + \sin^2\alpha_-\sin^22\theta \times T^{\mathrm{p}}P_{\mathrm{in}}P_{\mathrm{ref}}^{\mathrm{p}}},
    \label{eq:amp_cal}
\end{align}
where $P_{\mathrm{in}}$ and $P_{\mathrm{ref}}^{j}$ are the incident beam power onto the sample and the reference beam power on each PD, respectively.
As long as the reference beam power on each PD is characterized by the power meter (S132C from Thorlabs), Eq. (\ref{eq:amp_cal}) can be simplified as
\begin{align}
    \frac{\left(A^{\mathrm{p}}\right)^2/\left(T^{\mathrm{p}}P_{\mathrm{ref}}^{\mathrm{p}}\right)}{\left(A^{\mathrm{s}}\right)^2/\left(T^{\mathrm{s}}P_{\mathrm{ref}}^{\mathrm{s}}\right) + \left(A^{\mathrm{p}}\right)^2/\left(T^{\mathrm{p}}P_{\mathrm{ref}}^{\mathrm{p}}\right)}
    = \frac{\sin^2\alpha_-\sin^22\theta}{\left(1-\sin^2\alpha_-\sin^22\theta\right) + \sin^2\alpha_-\sin^22\theta}
    = \sin^2\alpha_-\sin^22\theta.
    \label{eq:amp_cal2}
\end{align}
Since $A^j$, $T^j$ and $P_{\mathrm{ref}}^{j}$ are measured values, we can derive $\sin^2\alpha_-\sin2\theta$ using Eq. (\ref{eq:amp_cal2}).


\subsubsection{Phase}
In order to measure the birefringence, $\alpha_-$, the phase offset, $\phi_{\mathrm{offset}}$, needs to be properly characterized.
We took the phase offset map prior to the birefringence measurement to compensate the phase offset.
In our demonstration setup, this can be done by setting the orientation of the QWP fast or slow axis to $45^\circ$.
From Eq.(\ref{eq:phase_diff}), when the slow axis ($\alpha_->0)$ is oriented to $\theta=45^\circ$, $\phi_{\mathrm{diff}}$ becomes $-\pi/2 + \phi_{\mathrm{offset}}$ so that the phase offset can be evaluated.

\subsection{Birefringence distribution}

\begin{figure}[htbp]
\centering\includegraphics[width=12.5cm]{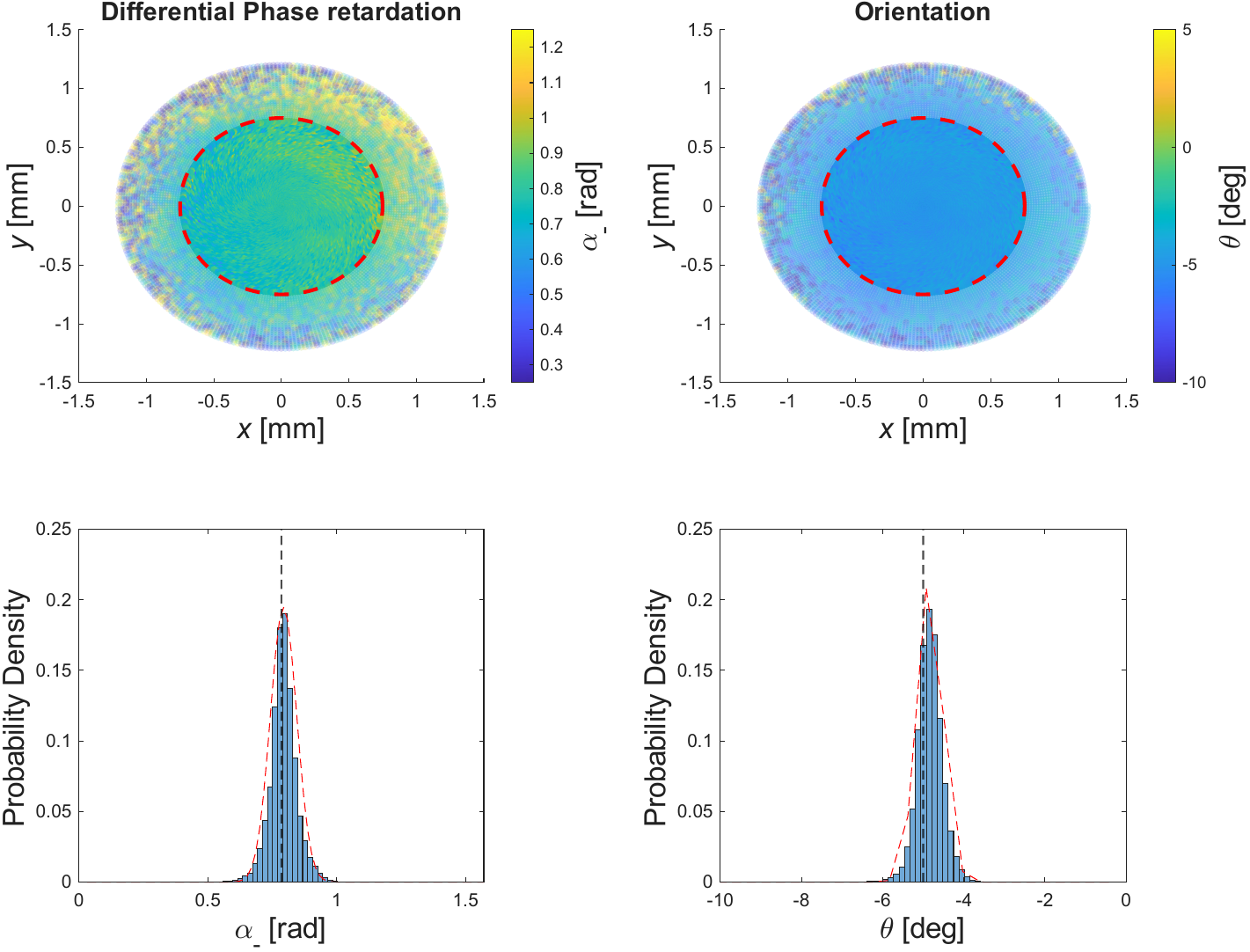}
\caption{
2D birefringence distribution maps and their histograms.
Red dashed circles in distribution maps describe the test beam diameters at the PDs.
Red dashed lines in histograms are the fitted curves, and black dashed lines are the expected values.
The distributions within the beam diameters are highlighted.
}
\label{Fig:map}
\end{figure}

To assess the validity and repeatability of the proposed heterodyne method, birefringence maps of the QWP were measured.
Five sets of measurements were taken for each of the three different orientations ($|\theta|\approx5^{\circ},\,11^{\circ},\,40^{\circ}$).
Fig. \ref{Fig:map} shows the typical results of 2D birefringence distributions, $\alpha_-$ and $\theta$, and their probability densities within the beam diameter.
This time, the slow axis of the QWP sample ($\alpha_->0$) was set to $\theta=-5^\circ$.
With this configuration, the test beam power on each PD becomes about $12\unit{mW}$ and $250\unit{\mu W}$ on s- and p-pol PDs, respectively.
The reference beam power was adjusted to $P_{\mathrm{ref}}^{\mathrm{s}} = 4.12\unit{mW}$ and $P_{\mathrm{ref}}^{\mathrm{p}}=18.6\unit{mW}$.
Note that for other axis orientations, i.e., $|\theta|\approx11^{\circ},\,40^{\circ}$, we set the reference beam power on each PD as $P_{\mathrm{ref}}^{\mathrm{s}} = P_{\mathrm{ref}}^{\mathrm{p}} = 11.4\unit{mW}$.

The red dashed circles in distribution maps (upper two plots) are the beam diameters on the PDs.
Within the beam diameter, birefringence distributions ($\alpha_-$ and $\theta$) are well reconstructed and their probability density distributions are well fitted by Gaussian distribution.
Red dashed curves in the histograms represent the fitted results.
The mean values and standard deviations obtained from these fittings are $(0.79\pm0.05)\unit{rad}$ and $-4.83^\circ\pm0.31^\circ$ for $\alpha_-$ and $\theta$, respectively.
Table \ref{tab:results} summarizes the obtained results.

Note that when we include the probability distributions outside the beam diameters, they largely deviate from Gaussian distribution.
In particular, their kurtoses are larger than 9.
As can be seen in Fig. \ref{Fig:map}, the values around the edge are deviated from the mean value largely.
Nevertheless, most values are distributed around the mean value within $\pm0.05\unit{rad}$, and this method is considered to be valid for estimating the birefringence distribution.





\begin{table}[htbp]
\caption{
Summary of the results.
$\alpha_{-,\,{\mathrm{meas}}}$ and $\theta_{\mathrm{meas}}$ are the measured values.
$|\alpha_{-,\,{\mathrm{approx}}}|$ are the estimated values using Eq. (\ref{eq:approx2}).
The orientations shown in this table are set by rotating the QWP.
Note that when the axis orientation is $\approx40^\circ$, due to the systematic error in $\alpha_{-,\,{\mathrm{meas}}}$, the calculated $\theta_{\mathrm{meas}}$ becomes $45^\circ$ as mentioned in Sec. \ref{sec:extraction}.
This corresponds to the white region shown in the right plot in Fig. \ref{Fig:OpticalLoss}.
}
\label{tab:results}
\centering
\begin{tabular}{ccccc}
\hline
$\alpha_{-,\,{\mathrm{meas}}}$ & $\theta_{\mathrm{meas}}$ & $|\alpha_{-,\,{\mathrm{approx}}}|$ & orientation & axis  \\
\hline
$(0.794\pm0.051)\unit{rad}$ & $-4.83^\circ\pm0.31^\circ$ & $0.787\unit{rad}$ & $\approx-5^\circ$ & slow ($\alpha_->0$)  \\
$(0.796\pm0.049)\unit{rad}$ & $-4.85^\circ\pm0.29^\circ$ &  \\
$(0.784\pm0.048)\unit{rad}$ & $-4.86^\circ\pm0.28^\circ$ & \\
$(0.784\pm0.048)\unit{rad}$ & $-4.86^\circ\pm0.28^\circ$ &  \\
$(0.793\pm0.048)\unit{rad}$ & $-4.86^\circ\pm0.30^\circ$ &  \\
\hline
$(-0.803\pm0.047)\unit{rad}$ & $10.56^\circ\pm0.74^\circ$ & $0.793
\unit{rad}$ & $\approx11^\circ$ & fast ($\alpha_-<0$) \\
$(-0.796\pm0.049)\unit{rad}$ & $10.64^\circ\pm0.77^\circ$ &  \\
$(-0.796\pm0.049)\unit{rad}$ & $10.62^\circ\pm0.77^\circ$ &  \\
$(-0.796\pm0.049)\unit{rad}$ & $10.51^\circ\pm0.76^\circ$ & \\
$(-0.796\pm0.049)\unit{rad}$ & $10.53^\circ\pm0.78^\circ$ & \\
\hline
$(0.701\pm0.020)\unit{rad}$ & $45^\circ$ & $0.717
\unit{rad}$ & $\approx40^\circ$ & slow ($\alpha_->0$) \\
$(0.702\pm0.022)\unit{rad}$ & $45^\circ$ & \\
$(0.702\pm0.022)\unit{rad}$ & $45^\circ$ &  \\
$(0.705\pm0.024)\unit{rad}$ & $45^\circ$ &  \\
$(0.701\pm0.016)\unit{rad}$ & $45^\circ$ &  \\
\hline
\end{tabular}
\end{table}

\section{Discussion}

The presented birefringence measurement method works well with the tabletop PPC setup.
In this section, we discuss a possible setup in an actual GWD, and its detectable limit.
Also, we discuss future plans toward implementations and a comprehensive understanding of how birefringence affects the performance of GWDs.

\subsection{Uncertainty}

As shown in Table \ref{tab:results}, the standard deviations in measured birefringence are about $6\%$ when the axis orientation is small, while it is reduced to about $2.5\%$ when the orientation becomes large.
This is considered to be the differences in the signal-to-noise (S/N) ratio for the different axis orientation.
For instance, when $\theta\approx11^\circ$, the test beam power on the p-pol PD is about $1\unit{mW}$.
On the other hand, that becomes about $10\unit{mW}$ when $\theta\approx40^\circ$.
Since the beatnote signal is proportional to the square root of the beam power, the S/N ratio at $\theta\approx40^\circ$ improves by about a factor of 3.
This results in the smaller uncertainties in measured values when $\theta\approx40^\circ$.

In our setup, one of the most dominant noise sources is the quantization noise, which is often called ADC noise.
This noise is introduced by the ADCs, which convert analog signals into digital signals, and its noise floor is about $4\times10^{-6}\unit{V/\sqrt{Hz}}$.
For practical implementation in a GWD, a high-end field programmable gate array (FPGA) is used as a data acquisition system \cite{Schaaf2016, Agatsuma2019}.
In Ref. \cite{Agatsuma2019}, the ADC noise floor of such high-end FPGA is estimated to be $\sim1.3\times10^{-8}\unit{V/\sqrt{Hz}}$.
Therefore, once the FPGA-based data acquisition system is implemented, the birefringence measurement accuracy can be improved.
This is one of the future works toward implementing the PPC for a GWD.

\subsection{Implications to GWDs}


\begin{figure}[htbp]
\centering\includegraphics[width=10cm]{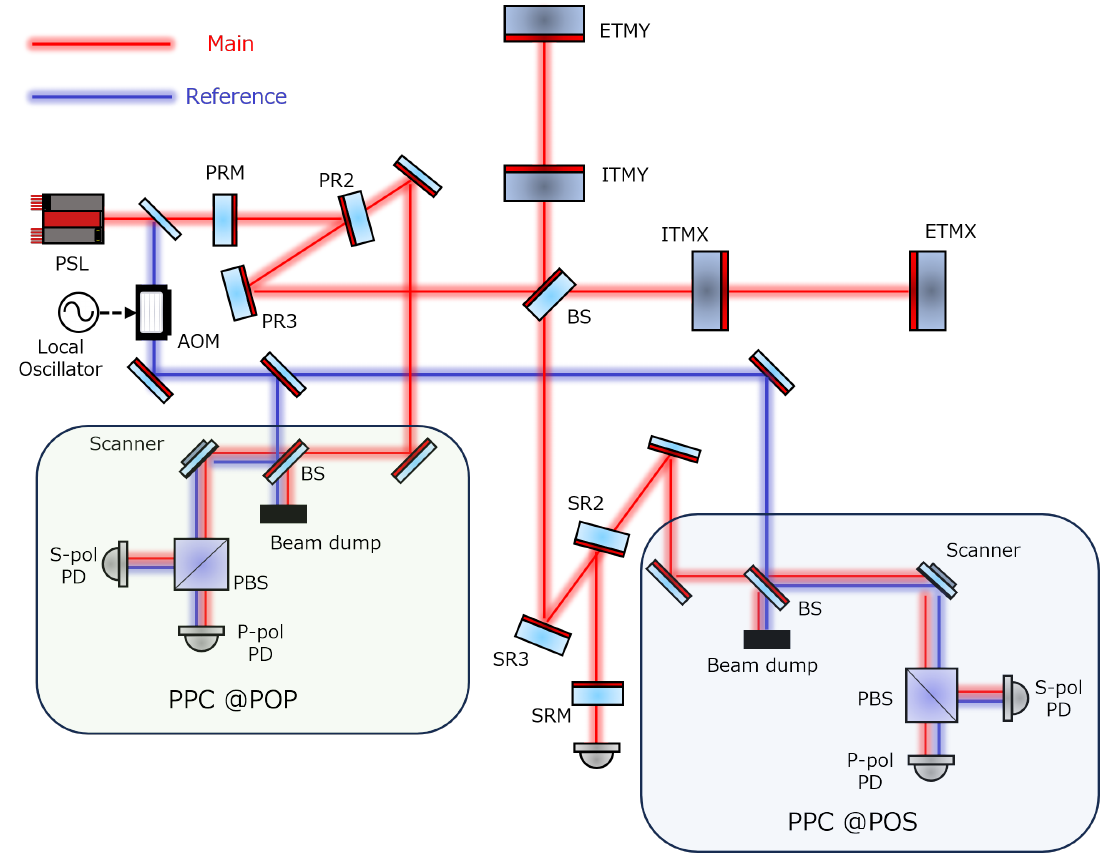}
\caption{
Simplified schematic of the possible \textit{in situ} birefringence characterization setup in a gravitational wave detector.
Part of the main beam is picked off before entering the interferometer and its frequency is shifted by the AOM in order to use as the reference beam for the PPC.
}
\label{Fig:PPCGWD}
\end{figure}

Fig. \ref{Fig:PPCGWD} shows a possible \textit{in situ} PPC setup in a GWD.
In this figure, we consider the case that we have two detection tables as can be seen in KAGRA, called POP and POS, where we pick off the main beam from the interferometer \cite{Akutsu2021}.
Since these picked off beams are affected by the test mass birefringence, we are able to investigate the polarization state of the main beam, hence birefringence, by using the PPC.

As reported in Ref. \cite{Michimura_note}, the BS introduces a relative phase delay between s- and p-polarized beams due to the non-zero angle of incidence, hence the phase offset, which corresponds to $\phi_{\mathrm{offset}}$ in Eq. (\ref{eq:phase_diff}).
In the same way, the NPBS2 shown in Fig. \ref{Fig:layout}, which combines the main and reference beams, will induce the relative phase delay.
Steering mirrors, which pick off the main beam from the interferometer to the detection table, also can contribute to the phase offset because of their non-zero angles of incidence.
As long as the phase offset ($\phi_{\mathrm{offset}}$) induced by those optics is pre-characterized during the detector commissioning period, birefringence distribution can be measured.
Note that future GWDs may install the folding mirrors in the power recycling cavity between the input test mass and the BS as shown in Refs. \cite{Rowlinson2021}.
In that case, the beam may be picked off from the folding mirror closest to the ITM, which will mitigate the impact of the relative phase delay in the BS.
Additionally, as long as the angles of incidence onto the folding mirror and steering mirrors are kept less than $\sim5^{\circ}$, the phase offset may be kept sufficiently small, typically less than $\sim10^{-3}\unit{rad}$ \cite{Xiao2020}.
With this optical configuration, the phase offset ($\phi_{\mathrm{offset}}$) might be negligible.
Toward implementation in future GWDs, further studies will be needed to minimize the phase offset and establish the calibration method.


\subsection{Detectability}

We discuss the detectability of substrate birefringence in future GWDs.
For the case of Cosmic Explorer, the circulating beam power inside the PRC will become about $10\unit{kW}$.
Typically, the folding mirror has the transmissivity of $T\sim200\unit{ppm}$ \cite{Aasi2015}.
This will allow us to pick off a total beam power of
\begin{align}
    \frac{1}{2}P_{\mathrm{BS}}T &\sim1\unit{W}.
\end{align}
Our PPC setup can detect birefringence even with $\sim250\unit{\mu W}$ beam power on the PD.
On the other hand, the main polarization beam has relatively high power ($\sim1\unit{W}$), which is not suited for the PDs used in the PPC.
Therefore, the main polarization beam power needs to be properly attenuated while the other polarized beam power is kept constant.
As long as the main polarization beam power is reduced to $\sim10\unit{mW}$ without affecting the beam power of the other polarization, the detectable optical loss due to the birefringence, $\mathscr{L}$, becomes
\begin{align}
    \mathscr{L} = \frac{250\unit{\mu W}\times2}{1\unit{W}} = 0.05\%.
    \label{eq:loss_limit}
\end{align}
Note that the factor of $2$ in the above equation takes into account of the effect of the combining BS in the PPC (NPBS2).
Further improvement may increase the detectability of birefringence. 
However, the detectability may ultimately be limited by the polarization extinction ratio of the PBS which splits the beam into two PDs (PBS2 in Fig. \ref{Fig:layout}).



\begin{figure}[htbp]
\centering\includegraphics[width=11cm]{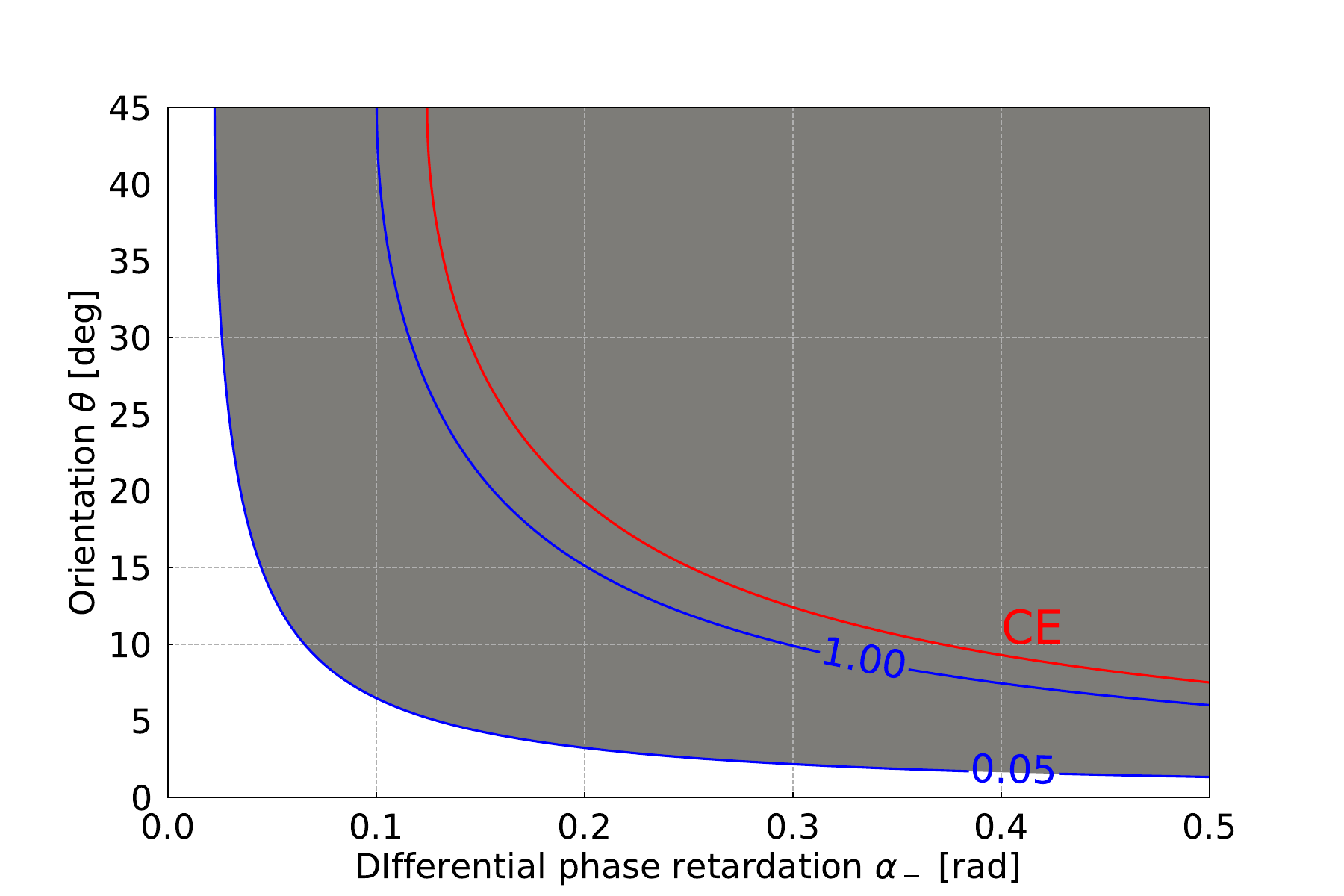}
\caption{
Detectability of the substrate birefringence in future GWDs.
The PPC can detect the substrate birefringence withing the region colored gray.
The boundary corresponds to the birefringent optical loss of $0.05\%$.
The red curve corresponds to the requirement to maintain the PRG of 65 for CE.
Also, the curve whose optical loss corresponds to $1\%$ is plotted as a reference.
}
\label{Fig:limit}
\end{figure}

Here, we compare the PPC detectability and required sensitivity to probe the optical loss value that must be satisfied to achieve the designed PRG ($G_{\mathrm{PRG}}$).
According to Refs. \cite{Winkler1994, Krüger2016}, the optical loss induced by birefringence ($L_{\mathrm{loss}}$), needs to satisfy
\begin{align}
    L_{\mathrm{loss}} < \frac{1}{G}.
\end{align}
For the case of Advanced LIGO detectors, they achieved PRG of $50$ during the 4th observing run \cite{Capote2025}.
To maintain this PRG level, the birefringence optical loss needs to be $L_{\mathrm{loss}}<2\%$.
Cosmic Explorer is planning to maintain the PRG of $65$ \cite{Hall2021}, yielding the birefringence optical loss less than $1.5\%$.
From Eq. (\ref{eq:loss_limit}) and Fig. \ref{Fig:limit}, the PPC setup demonstrated in this paper will have enough sensitivity to detect the birefringence optical loss which can reduce the PRG.

Fig. \ref{Fig:limit} shows the possible detectable limit of the PPC and the requirement to maintain the PRG of the GWDs discussed above.
The current PPC setup already has enough sensitivity to determine whether birefringence is affecting the PRGs of the Advanced LIGO detectors and Cosmic Explorer, and may probe the birefringence of the substrate ($\alpha_-$) down to $\sim0.05\unit{rad}$.

\subsection{Future plans}

As mentioned in Sec. IV of Ref. \cite{Michimura2024}, when the KAGRA arm cavity is locked on resonance, the power loss to the orthogonal polarization is reduced.
This is considered to be due to the Lawrence (or conjugation) effect \cite{Lawrence_thesis, Michimura2024}.
However, this reduction of the power loss may not be valid for sidebands.
While the carrier field is a superposition of the prompt reflection and the leakage fields of the cavity, the sideband fields are only prompt reflection fields.
We are planning to develop an optical cavity system and introduce sidebands using an electro-optic modulator to investigate the Lawrence effect by using the PPC.
Also, further improvement of the S/N ratio is going to be performed.
This future work will pave a way to understanding how substrate birefringence affects the interferometer condition. 

\section{Conclusion}

We presented the heterodyne polarimetry method which enables \textit{in-situ} birefringence characterization of a gravitational wave detector.
To verify the proposed birefringence mapping method, we have developed the PPC setup, and demonstrated the proof-of-concept using the QWP.
The measured birefringence agreed with the predicted value and the standard deviation in each measurement was typically less than $6\%$.
Since the PPC is based on the scanning pin-hole phase camera used in the GWD, it can be applied to gravitational wave detectors and can diagnose birefringence conditions inside the interferometers.
Especially, when the PPC is used in GWDs which will employ high power laser beams such as Cosmic Explorer, it can probe substrate birefringence with enough sensitivity to determine whether it affects the PRG.


\begin{backmatter}
\bmsection{Funding}
The Royal Society's Research Grant, RGS\textbackslash R2\textbackslash212142.

\bmsection{Acknowledgment}
The authors greatly thank Masayuki Nakano, Hiroaki Yamamoto, Yoichi Aso, Haoyu Wang, Carl Blair, Evan Hall, and the LIGO Scientific Collaboration Advanced Interferometer Configuration working group for useful discussions and feedback. 
This paper has LIGO Document number LIGO-P2500756.

\bmsection{Disclosures}
The authors declare no conflicts of interest.

\bmsection{Data availability} Data underlying the results presented in this paper are not publicly available at this time but may be obtained from the authors upon reasonable request.

\end{backmatter}


\bibliography{reference}






\end{document}